\documentclass[review]{elsarticle}

\usepackage{lineno,hyperref}
\modulolinenumbers[5]

\journal{Current Opinion in Structural Biology}

\begin{document}

\begin{frontmatter}

\title{Frustration, function and folding}

%% Group authors per affiliation:
\author{Diego U. Ferreiro}
\address{Protein Physiology Lab, FCEyN-Universidad de Buenos Aires. IQUIBICEN/CONICET. Intendente GŸiraldes 2160 - Ciudad Universitaria - C1428EGA, Buenos Aires, Argentina. Corresponding author email: ferreiro@qb.fcen.uba.ar}

\author{Elizabeth A. Komives}
\address{Department of Chemistry and Biochemistry, University of California, San Diego, 9500 Gilman Drive, La Jolla, CA 92092-0378, USA.}

\author{Peter G. Wolynes}
\address{Center for Theoretical Biological Physics, Department of Chemistry, Department of Physics, Department of Biosciences, 6100 Main Street, Houston, TX 77005, USA.}

%% or include affiliations in footnotes:
%\author[mymainaddress,mysecondaryaddress]{Elsevier Inc}
%\ead[url]{www.elsevier.com}

%\author[mysecondaryaddress]{Global Customer Service\corref{mycorrespondingauthor}}
%\cortext[mycorrespondingauthor]{Corresponding author}
%\ead{support@elsevier.com}

%\address[mymainaddress]{1600 John F Kennedy Boulevard, Philadelphia}
%\address[mysecondaryaddress]{360 Park Avenue South, New York}

\begin{abstract}

Natural protein molecules are exceptional polymers. Encoded in apparently random strings of amino-acids, these objects perform clear physical tasks that are rare to find by simple chance. Accurate folding, specific binding, powerful catalysis, are examples of basic chemical activities that the great majority of polypeptides do not display, and are thought to be the outcome of the natural history of proteins. Function, a concept genuine to Biology, is at the core of evolution and often conflicts with the physical constraints. Locating the frustration between discrepant goals in a recurrent system leads to fundamental insights about the chances and necessities that shape the encoding of biological information.

\end{abstract}

%\begin{keyword}
%\texttt{elsarticle.cls}\sep \LaTeX\sep Elsevier \sep template
%\MSC[2010] 00-01\sep  99-00
%\end{keyword}
%
\end{frontmatter}

%\linenumbers

\paragraph{Highlights} 
\begin{itemize}

\item Protein functional signals conflict with robust folding
\item Local frustration sculpts protein dynamics
\item Functionality emerges through frustration

\end{itemize}

\section{Main text}
\paragraph{Introduction} 
    Inorganic crystals are beautiful like wallpaper- every design element fits repetitively in place. Biomolecules are beautiful too but, like a painting by a great master, are made up of diverse parts, each still falling in place, where small details of how they are put together suggest action or life. Repetition is satisfying but action requires some conflict or frustration. Frustration occurs when a physical system is unable simultaneously to achieve minimum energy individually for each and every part of it \cite{0022-3719-10-18-008}. Frustration can happen for geometric reasons (as in a triangular spin lattice or a complicated protein topology) and/or due to competition between the interactions of the basic elements. The application of this concept to protein molecules \cite{Bryngelson:1987zr} paved the way to the development of the Energy Landscape Theory of protein folding, which provides powerful tools for understanding natural protein molecules \cite{Wolynes:2015uq, Wei:2016vn}. The basic notion is the recognition that natural proteins are evolved polymers distinguishing them from random polypeptides thrown together entirely by chance. It is at the protein level that conflicting biological goals meet in the specification of the sequences. In order to fold robustly, proteins must satisfy a large number of local interactions simultaneously, a task that is feasible when frustration between interactions of the elements is low \cite{Bryngelson:1987zr, Tzul:2017kx}. Beyond folding, however proteins perform chemical activities that impose further restrictions on the sequences that encode a given fold, possibly conflicting with the necessity of self-organization \cite{Lubchenko:2008kh}. Looking for the deviations of the expectations of structural stability hints at other teleonomical goals that are needed for action (Figure 1).
    
To see the conflicts encoded in protein sequence and structure one needs a reliable way for measuring the degree of satisfaction of a general energy function, a daunting task for large molecular objects such as proteins, where thousands of atoms interact by a myriad of weak forces. It is apparent, however, that we do not need to get to the fundamental quantum mechanics, as many of the forces can be accounted for in coarse-grained descriptions. Useful approximations to the energy functions are now at hand, and are being developed at different levels. Full atomic force fields are accurate enough to analyze the folding of small proteins (albeit at large computational cost), and multiple heuristics have led to ways to design sequences that fold to simple topologies \cite{Parmeggiani:2015qf, Woolfson:2015dq, Huang:2016nx}. These approximations rely on the fact that many of the interactions can be modeled with effective forces averaged over the solvent environment, such that the polymer can be described as being made of pseudoatoms that encode distinct properties \cite{Schafer:2014ve, Capelli:2014cr, Gopi:2017ve}

Having a reliable way of measuring the overall free energy of a protein structure, one can explore how the free energy varies when the sequence or the structure of the protein changes. Ten years ago a simple heuristic method to explore these relations was presented \cite{Ferreiro:2007ys, Parra:2016fk}. To analyze the existence of energetic conflicts in a folded protein, the energy of structural or sequence decoys is measured with respect to the native state. A local frustration index is defined as the Z-score of the free energy of parts of the native structure with respect to the distribution of the energy of rearranged decoys. If a native pair of interacting residues has an energy that lies in the most favorable end of the distribution, the interaction is labelled as Ôminimally frustratedÕ, as most changes in sequence in that location will destabilize the overall structure. In general, about 40\% of the native contacts found in natural globular domains fall into this class, in line with the theoretical expectations and experimental results \cite{Ferreiro:2007ys}. About half of the interactions can be labelled as ÔneutralÕ as they do not contribute distinctively to the total energy, and around 10\% of the interactions are Ôhighly frustratedÕ. These are regions in which most local sequence or structural changes would lower the free energy of the system. These frustrated regions are typically found as patches on the surface of globular domains. They must be held there over evolutionary time as well as physiological time at the expense of other interactions, that is, they conflict with the robust folding of a domain. The adaptive value for a molecule to tolerate spatially localized frustration arises from the way such frustration sculpts protein dynamics for specific functions. In a monomeric protein the alternate configurations caused by locally frustrating an otherwise largely unfrustrated structure provide specific control of the thermal motions guiding them in useful directions \cite{Ferreiro:2011kl, Li:2011hc, Fuglestad:2013ij, Munshi:2015zr, Souza:2017fk}. Alternatively a site that is frustrated in a monomeric protein may become less frustrated in the final larger assembly of this protein with partners, thus guiding specific association \cite{Ferreiro:2007ys, Zheng:2013ly, Matsushita:2013oq}. We will review here recent findings insights that come from analyzing the specifics of local frustration in several systems. For a detailed discussion of the basics of frustration biophysics the reader is referred to \cite{Ferreiro:2014kx}.

\paragraph{Topology can be frustrating} 
Frustration can be reflected in the topology of a proteinsÕ native states. Assuming that no energetic conflicts are present in a folded molecule, the chain connectivity by itself strongly restricts the sampling of the conformational space. Forming structure in one region may hinder the consolidation of structure in a distant part. Such Ôtopological frustrationÕ can be quantified using folding simulations with structure-based models \cite{Giri-Rao:2016bs}. Gosavi et al have shown how subtle differences in the native topology can give rise to large biological effects was shown in the Interleukin (IL) family. Despite having similar three-dimensional structures and stabilities, IL-1$\beta$ promotes downstream signaling, whereas IL-1Ra inhibits it. The folding traps caused by topology that distinguish these proteins make  IL-1$\beta$ fold more slowly than IL-1Ra. The differences in the landscape can be ascribed mainly in two loops, which when mutated can switch the functional forms \cite{Gosavi:2008fv}. Regions that participate in function are inferred to cause the different folding traps. 

The flavodoxin-like fold is an old protein architecture. Proteins with this fold often misfold in search of their functionally active forms. This susceptibility to misfold is caused by the timing of the consolidation of structure, forming partially folded intermediates \cite{Houwman:2017dz}. The differential stabilization of the intermediates can be related to the resistance to aggregation and degradation which change function in a system sense. Sequence permutation is a straightforward way to probe the effect of topological frustration. Nobrega et al analyzed the effect of circular permutants of CheY, a flavodoxin like bacterial protein. They showed that the stability of the folding trap can be modulated by changing the location of the termini, modulating the structures and stabilities of the kinetic traps that appear early in folding of CheY \cite{Nobrega:2014fu}. It is intriguing to note that only a few natural circular permutants of protein folds have been found. It remains to be determined if this is a reflection of an intrinsic physical constraint where some permutants are much more topologically frustrated than others, perhaps impeding folding \textit{in vivo}. 

Frustration in the early stages of folding is not always purely topological, it can also come about by the energetic differences given by sequence specification. DiSilvio et al have recently analyzed two alternative splice forms of a PDZ domain that share a nearly identical sequence and structure \cite{Di-Silvio:2015kl}. The kinetic characterization of site-directed mutants reveals that the late stages for folding are very robust and biased by native topology, but the early stages are more malleable and dominated by energetic local frustration.

    Some of the general aspects of frustration in folding can be studied simply by adding attractive non-native interactions to structure-based models, allowing for a competition with native contacts \cite{Oakley:2011uq}. Simple theoretical arguments show that including weak frustrating interactions does not always disturb but actually may facilitate folding \cite{Clementi:2004qa}. In the same line, Contessotto et al showed that the effect of homogeneous energetic frustration can play off topological frustration effects. Energetic frustration effects are stronger in the more intricate topologies \cite{Contessoto:2013mi}. Attractive non-native interactions help collapse the protein chain and thus entropically facilitate the search of native interactions. 
    
    \paragraph{Frustration and interaction} 
    Proteins seldom act alone. The prime activities of most proteins are related to the formation of higher order structures by specific binding between domains. Strong interactions are brought about by complementarity of the surfaces by means of chemical interactions that are essentially the same types as those that stabilize the folds. Sequence modifications that would promote binding then conflict often with the folding of one of the partners, for example when a large interaction hydrophobic surface is exposed. Local frustration calculations performed on the unbound forms of heterodimeric complexes indeed show that the residues in the protein-protein interfaces are enriched in highly frustrated interactions \cite{Ferreiro:2007ys}. If the binding surfaces are too big with respect to the area of the domains, they compromise the folding of the unbound domains. It is interesting to note that coupling between folding and binding is more prevalent when the interaction surface area is large \cite{Zhuravlev:2010pi}. The transient interaction between ferredoxin and photosystem I was analyzed bringing together models of the structures of the individual components, biochemical data and molecular docking. Two regions of high local frustration were identified on the surface of ferredoxin predicting that these regions interact predominantly with regions of high frustration on photosystem I molecule, including several residues pinpointed by other experimental studies \cite{Cashman:2014ff}.
Frustration in binding can kick in dynamically. Calmodulin (CaM) is a signaling protein that specifically recognizes and activates a variety of proteins. Exploring the molecular basis for target recognition it was found that two CaM target peptides, although similar in length and net charge, follow distinct routes that lead to a differential binding frustration \cite{Tripathi:2015lh}. The molecular origin of the binding frustration is caused by intermolecular contacts formed with one domain that need to be broken before intermolecular contacts form with the other domain. The appearance of binding kinetic traps determines the kinetics of the recognition process of proteins involving large structural fluctuations. 

    Conformational changes associated with binding exemplify how local frustration may modulate the interactions. Binding between TFIIIA and 5S RNA involves a large conformational change in three zinc fingers, involving the exposure and burial of several crucial DNA/RNA binding residues. Tan et al used an atomistic model of the specific recognition between finger 4/finger 6 and the 5S RNA and showed that one interaction introduces frustration into the nonspecific interactions between finger 5 and the 5S RNA, which then contributes to achieving the native binding specificity \cite{Tan:2013fu}. 

    The development of locally frustrated regions in a dynamic process allows the possibility for regulation. A striking example occurs in the interactions of the transcription factor NF$\kappa$B with DNA and I$\kappa$B, which involves the formation of either NF$\kappa$B-DNA or NF$\kappa$B-I$\kappa$B adducts as final species. I$\kappa$B$\alpha$ actively removes NF$\kappa$B from DNA target sites \cite{Potoyan:2016ye}, not by dissipating energy from an external source such as ATP, but rather by catalyzing the effectively irreversible formation of the NF$\kappa$B-I$\kappa$B complex. The mutually exclusive formation of these complexes involves free energy differences that originate from the frustrated electrostatic interactions between the negatively charged PEST region in I$\kappa$B and DNA \cite{Potoyan:2016qo}. The PEST occupies two different conformations in the NF$\kappa$B-I$\kappa$B$\alpha$ complex, one of which occupies the DNA-binding cavity. Indeed, neutralizing the negative charges in the I$\kappa$B$\alpha$ PEST sequence results in an NF$\kappa$B-I$\kappa$B complex that can still bind DNA \cite{Dembinski:2017fk}. In addition to the electrostatic frustration of the DNA cavity by the I$\kappa$B$\alpha$ PEST sequence, specific interactions of I$\kappa$B with the NF$\kappa$B dimerization domains draws the DNA-binding domains closer together hindering reassociation of the DNA and consolidating the NF$\kappa$B-I$\kappa$B complex.
    
    \paragraph{Frustration in natural variants} 
    Analysis of the local frustration patterns in various members of a protein family allows the investigation of the invariant aspects of the energy contributions. Such an analysis was performed on the ankyrin repeat family, systems for which structural data are abundant \cite{Parra:2015tw}. The family displays high internal symmetry. Frustration analysis shows there is a common core of minimally frustrated interactions in and between repeats that correlates with the thermodynamic stability of the repeat-array. In contrast, the highly frustrated patches in each protein are not conserved across the family but are tolerated in various regions in order to allow a wide range of functions \cite{Parra:2015tw}. The fact that synthetic proteins made from consensus sequences of this family are likely to fold can be explained by the linear correlation between the conservation of the energetic features in the repeat arrays and their sequence variations \cite{espada2017ff}. 
    
    Frustration is important in biomedicine. Sequence variations in the frustration patterns have been analyzed in a large dataset of rare single-nucleotide variants (SNVs) in coding regions of the human genome \cite{Kumar:2016il}. It was found that disease-related SNVs create stronger changes in localized frustration than do non-disease related variants, and that rare SNVs tend to disrupt local interactions to a greater extent than do common variants. Interestingly, somatic SNVs associated with oncogenes and tumor suppressor genes induce very different changes in the frustration patterns. Tumor suppressor mutations change the frustration more in the core than the surface by introducing loss-of-function events, whereas those associated with oncogenes manifest the opposite pattern, creating gain-of-function events sometimes by relieving frustration so as to interfere with the proper allosteric communication \cite{Kumar:2016il}. 
    
    Cell cycle progression depends on regulatory cyclin proteins that are frequently misregulated in human cancers. Structural analysis of the activating transitions of the CDK/cyclin complexes involve substantial structural changes, which were analyzed by integrating different levels of resolution for efficient sampling of the conformational space \cite{Floquet:2015zt}. The lowest-frequency normal modes described the transition between the open and closed conformations, and this transition is facilitated by distinct distributions of frustrated contacts in the complex. It was recently reviewed how the anisotropic nature of protein dynamics induces a protein response to external perturbations along a small number of intrinsic large-amplitude directions \cite{Kitao:2017jl}. Changes in energetic frustration that occur along large-amplitude motions act as switches to regulate protein function and can be triggered by small external perturbations such as the binding of other molecules, leading to the emergence of allosteric control.
    
    Higher order aspects of frustration were investigated in frataxin, an iron binding protein that is involved with the assembly of ironÐsulfur clusters, and where mutations are associated with Friedreich's ataxia \cite{Roman:2012gb}. Careful reconstruction of the transition state ensembles of folding, informed by experimental data, revealed exposure of regions that are highly aggregation-prone \cite{Gianni:2014mb}. The regions that are relevant for binding are found partially misfolded in the transition state but these are resistant to aggregation. The competition between folding and function creates the possibility of misfolding. Preventing aggregation requires that the amino acid sequence to be optimized be highly resistant to aggregation, specifically in the regions involved in misfolding. 
        
    \paragraph{Conclusions} 
    There are some new and notable examples being described in which the signals that encode protein folding conflict with protein functionality. Stability-activity trade-offs have been described in many protein families \cite{Wiederstein:2005uq, Boucher:2016cq} reflecting the conflicting goals set by physiology. All these examples suggest that many extant protein sequences are actually quite close to the saturation of the coding capacity needed to simultaneously specify folding and function. Protein sequences, structures and activities can be quantified, yet the teleonomic, apparent purposefulness of biological function often involves several events coming together in a coordinated way \cite{MAYR:1961rq, monod1973hasard, Kauffman:2014fk}.  At any scale it is challenging to disentangle what are the meaningful parts of living things and what are the relevant interactions between them \cite{simondon2005individuation}. On top of the beauty that simple inorganic objects display, biomolecules and their assemblies encode information that sparkle our imagination about their relations, as these are always pieces of a larger body. At every level of biological organization we tend to seek a purpose for almost every detail of a natural contrivance, looking to distinguish meaning from noise, selection from drift, chance from necessity. Analyzing frustration is a tool in the search for meaning.

\begin{figure}[h]
    \centering
	\includegraphics[width=1 \textwidth]{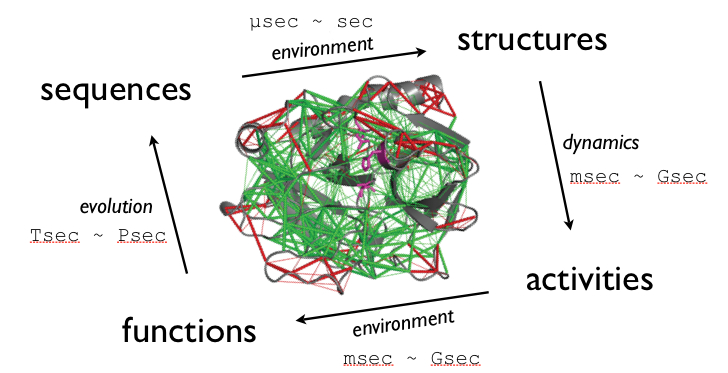}
	  \caption{The recurrent flow of biological information conflicts in natural proteins. In appropriate environmental conditions, amino acid sequences encode the formation of specific structures. Structures interconvert due to frustration and give rise to chemical activities, contributing to the specification of multiple biological functions. The functional structures restrict the exploration and fixation of the genomic sequences. These processes occur in timescales that span many orders of magnitude. Conflicts between folding and function can be located in extant proteins, reflecting functional constraints. At center a representation of local frustration on Thrombin, a modern protease \cite{Fuglestad:2013ij}. The backbone is shown as a continuous gray trace. The protein is networked by a connected set of minimally frustrated contacts (green) while there are patches of highly frustrated contacts (red). Thin lines represent water-mediated interactions and the catalytic residues are shown in magenta. Local frustration patterns were calculated with the frustratometer.tk server \cite{Parra:2016fk}}	
	\label{figure}
\end{figure}

\section{Acknowledgements}
We thank the members of our research groups for the enriching discussions, specially to Davit Potoyan, Weihua Zheng, Nick Schafer, Gonzalo Parra, Brenda Guzovsky, Roc\'io Espada, Nacho S\'anchez and Manolo Rodr\'iguez.

\section{Funding}

Funding: This work was supported by the Consejo de Investigaciones Cient\'ificas y T\'ecnicas (CONICET); the Agencia Nacional de Promoci\'on Cient\'ifica y Tecnol\'ogica [PICT2012/01647 to D.U.F.] and ECOS Sud - MINCyT n¡ A14E04. This work was supported by National Institute of General Medical Sciences [Grant  R01GM44557 to P.G.W and P01-GM071862 to E.A.K].  Additional support was provided by D. R. Bullard-Welch Chair at Rice University [Grant C-0016 to P.G.W.]

\section*{References}

\bibliography{mybibfile}

\section{Annotated References}
Outstanding

-ref14: An open web server tool to localize energetic frustration in proteins.

-ref40: Large scale analysis of local frustration effect in genomic variants associated to decease.

Special

-ref27: Kinetic characterization of site-directed mutants of two related proteins showing that the early stages of folding are dominated by local frustration.

-ref33: Molecular origin of the frustration caused in binding to CaM, determining the kinetics of the recognition process.

-ref36: Explores the functional motions of a transcription factor and its various complexes, showing that the function of the genetic switch is realized via an allosteric mechanism modulated by local frustration.

\end{document}